\documentclass[11pt,twoside]{article}
\usepackage{intlp}
\usepackage{amssymb,amstext,latexsym,xspace}
\usepackage{epsfig}

\newcommand{\ct}[1]{~\cite{#1}}
\newcommand{\lb}[1]{\label{#1}}
\newcommand{\Eq}[1]{~(\ref{#1})}
\renewcommand{\[}{\begin{eqnarray}}
\renewcommand{\]}{\end{eqnarray}}
\newcommand{\nn}{\nonumber}
\newcommand{\non}{\nonumber \\ }
\renewcommand{\=}{\equiv}

\newcommand{\Com}[2]{[#1\, ,#2]}

\newcommand{\ft}[2]{{\textstyle {\frac{#1}{#2}} }}

\newcommand{\ID}{\mbox{1\hspace{-.35em}1}}
\renewcommand{\i}{\mathrm{i}}

\newcommand{\cC}{\mathcal{C}}

\newcommand{\cE}{\mathcal{E}}
\newcommand{\cF}{\mathcal{F}}

\newcommand{\cN}{\mathcal{N}}
\newcommand{\cQ}{\mathcal{Q}}

\renewcommand{\i}{\mathrm{i}}
\newcommand{\gd}{\delta}

\newcommand{\Rn}{\ensuremath{\mathbb{R}}\xspace}

\newcommand{\2}{{{\rm L}^2 \big(\Rn^{29}\big) }}
\newcommand{\fg}{\mathfrak{g}\xspace}
\newcommand{\fk}{\mathfrak{k}\xspace}
\newcommand{\fe}{\mathfrak{e}\xspace}
\newcommand{\SLR}[1][2]{\ensuremath{SL(#1,\Rn)}\xspace}
\newcommand{\SpR}[1][2]{\ensuremath{Sp(#1,\Rn)}\xspace}
\newcommand{\SO}[1][{}]{\ensuremath{SO(#1)}\xspace}
\newcommand{\SOS}[1][{}]{\ensuremath{SO^{*}(#1)}\xspace}
\newcommand{\SU}[1][{}]{\ensuremath{SU(#1)}\xspace}

\newcommand{\6}[1][{}]{\ensuremath{E_{6 #1}}\xspace}
\newcommand{\7}[1][{}]{\ensuremath{E_{7 #1}}\xspace}

\newcommand{\E}{\ensuremath{E_{8(8)}}\xspace}
\newcommand{\EE}{\ensuremath{E_{7(7)}}\xspace}
\newcommand{\EEE}{\ensuremath{E_{6(6)}}\xspace}

\newcommand{\symfootnote}[2]
{\renewcommand{\thefootnote}{\fnsymbol{footnote}}\footnote[#1]{#2}\renewcommand{\thefootnote}{\arabic{footnote}}}

\def\moth{\mathsurround=0pt}
\newdimen\zo \zo=0pt

\def\tick{\leaders\hrule height 0.5ex depth 0pt \hskip 0.5pt}
\def\upboxfill{$\moth \setbox\zo\hbox{\tick}%
  \hskip 2pt\hbox to 0pt{$\tick$\hss}\hrulefill \hbox to 2pt{$\tick$\hss}$}
\def\underbox#1{\offinterlineskip{\mathord{\mathop{\vtop{\moth\ialign{##\crcr
      $\hfil\displaystyle{#1}\hfil$\crcr\noalign{}
      {\upboxfill}\crcr\noalign{}}}}\limits}}}
\def\dtick{\leaders\hrule height .34pt depth 0.5ex \hskip 0.5pt}
\def\downboxfill{$\moth \setbox\zo\hbox{\dtick}%
  \hskip 2pt\hbox to 0pt{$\dtick$\hss}\hrulefill%
  \hbox to 2pt{$\dtick$\hss}$}
\def\overbox#1{\mathop{\vbox{\moth\ialign{##\crcr\noalign{}
\downboxfill\crcr\noalign{\vskip 1pt\nointerlineskip}
      $\hfil\displaystyle{#1}\hfil$\crcr}}}\limits}

\newcommand{\undersym}[1]{\underbox{{}#1}}
\newcommand{\oversym}[1]{\!\overbox{{}#1}}
\begin{document}






\url{hep-th/0109005}

\title{The Minimal Unitary Representation of $E_{8(8)}$\symfootnote{2}{This
    work was supported in part by the NATO collaborative research grant
    CRG.~960188.}}

\author{M.~G\"unaydin${\,}^{a}$\symfootnote{3}{Work supported in part by the
    National Science Foundation under grant number PHY-0099548.},
    K.~Koepsell${\,}^{b}$, H.~Nicolai${\,}^{b}$}
\address{${}^{a}$ Penn State University,\\ Physics Department,\\
  University Park, PA 16802, USA}
\addressemail{murat@phys.psu.edu}

\address{${}^{b}$ Max-Planck-Institut f{\"u}r Gravitationsphysik,\\
  Albert-Einstein-Institut,\\ M\"uhlenberg 1, D-14476 Golm, Germany}
\addressemail{koepsell@aei.mpg.de, nicolai@aei.mpg.de}

\markboth{\it The Minimal Unitary Representation of $E_{8(8)}$}{\it
  M.~G\"unaydin et al.}


\begin{abstract}
We give a new construction of the minimal unitary representation
of the exceptional group \E on a Hilbert space of complex functions
in 29 variables. Due to their manifest covariance with
respect to the \EE subgroup of \E our formulas are simpler than previous
realizations, and thus well suited for applications in superstring
and M theory.
\end{abstract}

\vfill

\newpage

\section{Introduction}
In this paper, we present a new construction of the minimal
unitary representation of \E ({\it i.e.} the maximally split real
form of the exceptional Lie group $E_8$) on a Hilbert space of
complex functions in 29 variables\footnote{A preliminary account
of some of the results presented in section 2 has already appeared
in\ct{Koepsell}.}. The  minimal realizations of classical Lie
algebras and of $G_2$ were given  by Joseph a long time ago
\ct{Jose74,Jose76}. The existence of the minimal unitary
representation of \E was first proved by Vogan \cite{Vogan81} who
located it within the framework of Langland's  classification.
Later, the minimal unitary representations of  all simply laced
groups, including \E, \EE and \EEE  were constructed by Kazhdan
and Savin\ct{KazSav90}, and Brylinski and Kostant
\ct{BryKos94,BryKos94a,BryKos95,BryKos96} by rather different
methods. Gross and Wallach gave yet another construction of the
minimal representation of \E as well as for all exceptional groups
of real rank four \cite{GroWal94}. For the exceptional group
$E_{8(8)}$ the minimal orbit is 58 dimensional, corresponding to 
its quotient by a distinguished parabolic subgroup. The representations
of $E_{8(8)}$ over the spaces of functions on the 128 dimensional
coset space $E_{8(8)}/ SO(16)$ were  studied in \cite{ahv98}. 

While formulas for $G_2$ similar to the ones derived here for \E
can be found in\ct{BryKos94a}, however, an explicit realization of
the simple root (Chevalley) generators in terms of
pseudo-differential operators for the simply laced exceptional
groups was given only very recently\ct{KaPiWa01}, together with
the spherical vectors necessary for the construction of modular forms.
We present here an alternative realization of \E, which has the advantage
of yielding very compact formulas, in contradistinction to the rather
complicated expressions obtained by multiple commutation of the
simple root generators \cite{web}. The main reason for the
relative simplicity of our final expressions (\ref{e7gen}) and
(\ref{e7e8gen}) is their manifest $\EE$ covariance, in spite of
the fact that $\EE$ is not realized linearly. The compactness
of our formulas makes them especially suitable for applications in
string and M theory which have primarily motivated the present work.

Although the present article is mainly addressed to a physicists'
audience, let us first summarize the main results in a somewhat
more mathematical language. The minimal representation of a
noncompact group $G$ corresponds, in general, to the quantization
of its smallest co-adjoint orbit. In the case of $E_{8(8)}$ the
minimal co-adjoint orbit is 58-dimensional. Starting from the
5-graded decomposition of the associated Lie algebra $\fe_8$  
given in eq.~(4) below, it can be obtained simply as the 
co-adjoint orbit of the highest root generator (designated by 
$E$ in eq.~(6)). In this 5-graded decomposition of $\fe_8$ one 
can readily identify a maximal non-semisimple subalgebra 
$\mathfrak{p}$  as the  annihilator of this generator, {\it i.e.}
\begin{equation}
[ \mathfrak{p}, E]=0
\end{equation}

The subalgebra  $\mathfrak{p}$ is generated by the grade zero
subalgebra $\fe_7$ together with the grade 1 and grade 2 elements
of the full $\fe_8$ Lie algebra ({\it i.e.} the generators 
$(E_{ij}, E^{ij})$ and $E$ in eq.~(6)). The orthogonal complement 
of this parabolic  subalgebra is a 58 dimensional 
nilpotent subalgebra $\mathfrak{n}$. In the notation of section~2,
the latter is generated by the grade 0 element $H$, the grade
$(-1)$ elements $(F_{ij}, F^{ij})$, and the grade $(-2)$ element
$F$. Acting with this nilpotent subalgebra on the generator $E$
corresponding to the highest root one obtains the generators
corresponding to the minimal orbit. Thus we see that the minimal orbit
can be identified with the coset space $E_{8(8)} / \mathfrak{P}$ where
$\mathfrak{P}$ is the parabolic subgroup generated by $\mathfrak{p}$.

We were led to this minimal unitary representation rather naturally
from the novel geometric realization of $E_{8(8)}$ on $\Rn^{57}$
found in our previous work \cite{GuKoNi00}. As shown there, the
exceptional groups can be realized via conformal or quasiconformal
transformations leaving invariant generalized ``light-cones''; these
transformations are analogous to the non-linear action of the
M\"obius group on the real line. Our construction relied essentially
on the connection with Jordan algebras and Freudenthal triple systems.
To proceed from there to the unitary realization on a suitable
(infinite dimensional) Hilbert space of functions, we must first
identify a phase space realization of this system and then quantize it.
For the \EE subgroup this phase space realization involves
28 coordinates $X^{ij}$ and 28 momenta $P_{ij}$, such that
all \EE variations can be realized via the canonical action
\[
\delta_\cQ X \equiv \{\cQ\, ,\,X\}_{\text{P.B.}} \; , \quad
\delta_\cQ P \equiv \{\cQ\, ,\,P\}_{\text{P.B.}} \label{Poisson}
\]
where $\cQ=\cQ(X,P)$ is the appropriate \EE charge. For the
canonical realization of the full \E we have to add the momentum
$p$ conjugate to the 57th coordinate $y$, and replace the M\"obius
action on $y$ by the symplectic realization acting on the two
dimensional phase space spanned by $y$ and $p$. The minimal
unitary representation of \E is then obtained by quantization,
i.e. replacement of the classical momenta by differential
operators. This requires in addition a prescription how to deal
with the ordering ambiguities arising in the non-linear
expressions for the Lie algebra generators. The latter 
are then realized as self-adjoint operators on some dense 
subspace of $\2$. This coordinate space (Schr\"odinger)
representation is reformulated in section~3 as an oscillator
realization in terms of annihilation and creation operators acting
in a particle basis. By going to the corresponding coherent state
basis of this oscillator realization labelled by 29 complex
coordinates one obtains  the Fock-Bargmann realization of the
minimal unitary representation over an Hilbert space of
holomorphic functions.

The irreducibility of the minimal representation is put in evidence
by showing that the quadratic Casimir operator of $E_{8(8)}$, when
expressed in terms of either the coordinate or the oscillator basis
reduces to a $c$-number; a general argument shows that likewise the
higher order Casimirs reduce to $c$-numbers in our realization.
When restricted to the subgroup $E_{7(7)} \times SL(2,\Rn)$ the minimal
representation decomposes into an infinite sum (actually an integral,
see \cite{BryKos94}) of irreducible representations of
$E_{7(7)} \times SL(2,\Rn)$. Remarkably, the quadratic Casimir of
$E_{7(7)}$ obtained from the minimal realization is identical with the 
quartic invariant of $E_{7(7)}$ when expressed as a function of the 28
coordinate and 28 momentum operators (or equivalently, the 28 annihilation
and 28 creation operators in the oscillator basis). The irreducible
representations of $SL(2,\Rn)$ that occur in the decomposition
of the minimal representation $E_{8(8)}$ w.r.t. $E_{7(7)} \times SL(2,\Rn)$
are all labelled by the eigenvalues of this quartic invariant.

As already mentioned our chief aim with the present work is to present
the results in such a way that they can be readily applied in the
context of string and M theory (for instance, readers familiar
with \cite{CreJul79} should have no difficulties
following our exposition). In particular, our results apply
in the context of the (super)conformal quantum mechanics description
of quantum black holes\ct{AlFuFu76,CDKKTP98,MicStr99}. In that work,
unitary representations of $SL(2,\Rn)$ played a crucial role in 
the classification of physical states. It is therefore an
obvious idea to extend these concepts to maximal supergravity.
Indeed, the candidate Hamiltonian which we obtain as one of the
\E Lie algebra generators reads
\[
L_0 = \ft14 \left[ p^2 + y^2 + 4\, y^{-2} I_4(X,P) \right]
\label{L0}
\]
which is precisely of the form studied by the authors
of\ct{AlFuFu76,CDKKTP98,MicStr99}, with the only difference that they
have a (coupling) constant instead of the differential operator
$I_4(X,P)$. Here $I_4$ is the quartic invariant of \EE expressed
as a function of the 28+28 variables $X^{ij}$ and $P_{ij}$; when
acting on one of the irreducible subrepresentations of \EE under
its subgroup $\EE \times SL(2,\Rn)$ it yields the associated
eigenvalue of the quadratic $\EE$ Casimir. As shown for instance
in\ct{GuKoNi00} in the context of the classical theory, the quartic 
invariant can assume both positive and negative values, and vanishes 
for $\ft12$ or $\ft14$ BPS black hole solutions of
$N=8$ supergravity\ct{KalKol96,FerMal98,FerGun98}. It is therefore
tempting to interpret (\ref{L0}) as the effective Hamiltonian
describing (in some approximation) $N=8$ quantum black holes, such
that every subrepresentation with a fixed eigenvalue of the operator
$I_4$ is identified with the space of physical states associated with
the corresponding black hole solution of $N=8$ supergravity. An
interpretation along similar lines has also been suggested in\ct{GibTow99}.
Interestingly, for the vanishing eigenvalues of $I_4$, the Hamiltonian
simplifies drastically, and the state space reduces to the well
known singleton representation of $SL(2,\Rn)$.

Another (and possibly related) physical application of minimal
representations has been outlined in\ct{KaPiWa01}. That work
evolved from an earlier attempt to determine the $R^4$ corrections
directly from the supermembrane and to understand them in terms of
so-called ``theta-correspondences''\ct{PNPW01}; see also\ct{SugVan01}
for a related attempt to come to grips with the non-linearities of
the supermembrane.

\mathversion{bold}
\section {Coordinate Space (Schr\"odinger) Representation}
\mathversion{normal}

We first recall some basic features of the non-linear realization of
\E on $\Rn^{57}$ coordinatized by 57 real variables $\{X^{ij},X_{ij},y\}$
($i,j=1,\ldots,8$), see\ct{GuKoNi00} for our notations and conventions,
and further details. A key ingredient in that construction
was the 5-graded decomposition of \E w.r.t.\ its subgroup
$E_{7(7)} \times \SLR[2]$. Denoting its Lie algebra by
$\fe_8$ we have
\[
\fe_8 &=&   \fg^{-2} \oplus \fg^{-1} \oplus \fg^0 \oplus \fg^{+1}
  \oplus \fg^{+2} \,.  \label{5grading}
\]
An important property of this decomposition is the fact that the
subspaces of grade $-1$ and $-2$ together form a maximal Heisenberg
subalgebra. The corresponding generators $E^{ij},E_{ij}\in\fg^{-1}$
and $E\in\fg^{-2}$ obey the commutation relations
\[\lb{heis-sub}
\Com{E^{ij}}{E_{kl}} &=& 2\i \,\delta^{ij}_{kl} E \,.
\]
Obviously, this algebra can be realized as a classical Poisson algebra
on a phase space with 28 coordinates $X^{ij}$ and 28 momenta
$P_{ij}\equiv X_{ij}$, and one extra real coordinate $y$ to
represent the central term. We have
\[
E^{ij} \;:=\; y\,X^{ij}\,, \quad
E_{ij} \;:=\; y\,P_{ij}\,, \quad E\;:=\;\ft12\,y^2.
\]
It is then straightforward to determine the generators of the
$\fe_7$ subalgebra in terms of these phase space variables (for
instance by requiring that they reproduce the \EE variations
{\it via} (\ref{Poisson})). They are realized by the following
63+70 bilinear expressions in $X^{ij}$ and $P_{ij}$
\[ \label{e7gen}
G^i{}_j  &:=& 2\, X^{ik}P_{kj}
              +\ft{1}{4} X^{kl}P_{kl} \,\gd^i_j \,,\non[2ex]
G^{ijkl} &:=& -\ft{1}{2} X^{[ij} X^{kl]}
              +\ft{1}{48} \epsilon^{ijklmnpq} P_{mn} P_{pq} \,.
\]

To extend this canonical realization to the full \E ,
we need one more variable, the momentum $p$ conjugate to $y$.
Combining the symplectic realization of $SL(2,\Rn)$ with the
non-linear variations (23) and (24) of\ct{GuKoNi00} we can then
deduce the classical (phase space) analogs of the \E generators.
Owing to the non-linearity of the realization, the resulting
expressions are not only quadratic, but go up to fourth
order in the phase space variables $X^{ij}$ and $P_{ij}$,
and moreover contain inverse powers of $y$. Thus all generators
of \E can be realized on a 58-dimensional phase space coordinatized
by $\{X^{ij},P_{ij}\,;\,y,p\}$. The minimal unitary 29 dimensional
representation of \E is then obtained by quantization,
i.e. by introducing the usual momentum operators
obeying the canonical commutation relations
\[
[X^{ij},P_{kl}] \;=\; \i\,\delta^{ij}_{kl}\,, \qquad
[y,p] \;=\; \i \,. \nn
\]
In elevating the \E generators to quantum operators the only
problem {\it vis-\`a-vis} the classical phase space description is
the non-commutativity of the coordinate and momentum operators,
which requires some ordering prescription (we note, that the \EE
generators (\ref{e7gen}) are insensitive to re-ordering the
coordinates and momenta). Since we are interested in finding a
unitary representation we insist that all operators are hermitean
w.r.t. to the standard scalar product on $\2$. We thus arrive at a
unitary representation of the \E Lie algebra in terms of
self-adjoint operators acting on (some dense subspace of) a
Hilbert space of complex functions in 29 real variables
$\{X^{ij},y\}$. We emphasize that this realization requires {\it
complex} functions, for the same reason that the Schr\"odinger
representation of the one-dimensional point particle requires
complex wave functions\footnote{ As we show in the next section
the generators of \E can be rewritten in terms of bosonic
annihilation and creation operators. In the corresponding
coherent state basis the Hilbert space will involve {\it holomorphic}
functions in 29 variables.}.

Before writing out the \E generators, let us list the \7[(7)]
commutation relations
\[
\Com{G^i{}_j}{G^{k\vphantom{l}}{}_l} &=&
  \i\,\gd{}_{j}^{k}\, G^i{}_{l} -\i\,\gd{}_{l}^{i}\, G^k{}_{j} \,,\non[2ex]
\Com{G^i{}_j}{G^{klmn}}  &=& -4\i\,\gd{}_{j}^{[k}\, G^{lmn]i}_{\phantom{k}}
                   -\ft{\i}{2}\, \gd{}^i_j G^{klmn}_{\phantom{j}}\,,\non[2ex]
\Com{G^{ijkl}}{G^{mnpq}} &=& \ft{\i}{36}\,\epsilon^{ijkls[mnp}\,
                                 G^{q]}{}_{s} \,.
\]
Observe that only the $SL(8,\Rn)$ subgroup acts by linear
transformations; its maximal compact subgroup $SO(8)$ is
generated by $G^-_{ij}$, where
\[
G^\pm_{ij} := \frac12 \left( G^i{}_j \pm G^j{}_i \right)\,.
\]
For later convenience, we also define
\[
G_{ijkl} &:=& \ft{1}{24} \epsilon_{ijklmnpq} G^{mnpq}\,,
\]
and the selfdual and anti-selfdual combinations
\[
G^\pm_{ijkl} := \frac12 \left( G^{ijkl} \pm G_{ijkl} \right)
\]
where the $G^-_{ijkl}$ are compact and the $G^+_{ijkl}$ non-compact.

The following generators extend this representation to the full \E
(we use the same notation for the elements of $\fe_8$ as in \cite{GuKoNi00},
in accordance with its 5-graded structure):
\[ \lb{e7e8gen}
E      &:=& \ft12\, y^2 \,,\non[2ex]
E^{ij} &:=& y\,X^{ij} \,,\non[2ex]
E_{ij} &:=& y\,P_{ij} \,,\non[2ex]
H      &:=& \ft{1}{2} (y\,p+p\,y) \,,\non[2ex]
F^{ij} &:=& -p\,X^{ij} + 2\i y^{-1}\, \Com{X^{ij}}{I_4(X,P)}\non
       &=&  -4\,y^{-1} X\oversym{^{ik}P_{kl}X^{lj}}
            - \ft{1}{2}\,y^{-1}(X^{ij}P_{kl}X^{kl}+X^{kl}P_{kl}X^{ij})\non
       &&
         +\ft{1}{12} y^{-1} \epsilon^{ijklmnpq} P_{kl}P_{mn}P_{pq}
         -p\,X^{ij}\,,\non[2ex]
F_{ij} &:=& -p\,P_{ij} + 2\i y^{-1}\, \Com{P_{ij}}{I_4(X,P)}\non
       &=&  4\,y^{-1} P\undersym{_{ik}X^{kl}P_{lj}}
            +\ft12\,y^{-1}(P_{ij} X^{kl}P_{kl}+P_{kl}X^{kl} P_{ij})\non
       &&
          -\ft{1}{12} y^{-1} \epsilon_{ijklmnpq} X^{kl}X^{mn}X^{pq}
          -p\,P_{ij}\,,\non[2ex]
F      &:=& \ft12 p^2 + 2 y^{-2} I_4(X,P)    \,.
\]
The hermiticity of all generators is manifest.
Here $I_4(X,P)$ is the fourth order differential operator
\[ \label{invgen}
I_4(X,P) &:=& -\ft12(X^{ij}P_{jk}X^{kl}P_{li}+P_{ij}X^{jk}P_{kl}X^{li})\non
         &&   +\ft18(X^{ij}P_{ij}X^{kl}P_{kl}+P_{ij}X^{ij}P_{kl}X^{kl})\non
         &&   -\ft{1}{96}\,\epsilon^{ijklmnpq} P_{ij}P_{kl}P_{mn}P_{pq} \non
         &&   -\ft{1}{96}\,\epsilon_{ijklmnpq} X^{ij}X^{kl}X^{mn}X^{pq}
               + \ft{547}{16} \,.
\]
As a function of $X^{ij}$ and $P_{ij}$ this operator represents the
quartic invariant of \EE because
\[ \label{quartic}
\Com{G^i{}_j}{I_4(X,P)} = \Com{G^{ijkl}}{I_4(X,P)} = 0 \,.
\]
We emphasize the importance of the ordering adopted in (\ref{invgen})
for the vanishing of (the second of) these commutators. Orderings that
differ from (\ref{invgen}) by a term containing the Euler operator
$i X^{ij} P_{ij}$ break \EE invariance. On the other hand, (\ref{quartic})
is insensitive to re-orderings which differ from (\ref{invgen}) only
by a $c$-number, and in the absence of a preferred ordering there
is thus no absolute significance to the additive constant appearing
in the definition of $I_4$. For instance, an admissible re-ordering is
\[\label{e7-casimir}
-\ft16\, G^i{}_j G^j{}_i -G^{ijkl} G_{ijkl}= I_4(X,P) - \ft{323}{16} \,.
\]
This relation confirms our previous assertion that, when acting on a given
representation of \EE, $I_4$ is just the quadratic \EE Casimir invariant,
up to an additive constant.

For the derivation of (\ref{e7e8gen}), we note that, as already pointed out
in\ct{BryKos94}, the crucial step is the determination of the operators
$E$ and $F$ corresponding to the lowest and highest root of \E, respectively.
The expressions for $F^{ij}$ and $F_{ij}$ then follow by commutation
with $E^{ij}$ and $E_{ij}$. While the derivation of\ct{BryKos94} relied
on a generalization of the so-called ``Capelli-identity'', the form
of our operators $E$ and $F$ follows directly from \EE invariance and
a scaling argument, up to the additive constant in (\ref{invgen}).
The latter originates from the re-ordering required to bring the
commutator of $F_{ij}$ and $F^{ij}$ into the ``standard form'' defined
by the r.h.s. of (\ref{invgen}).

As anticipated, all generators transform covariantly under the full \EE group.
\[
\Com{G^i{}_j}{E^{kl}} &=&  \i\,\gd{}^k_j\,E^{il} -\i\,\gd{}^l_j\,E^{ik}
                           -\ft{\i}{4}\gd{}^i_j\,E^{kl} \,,\non
\Com{G^i{}_j}{E_{kl}} &=&  \i\,\gd{}^i_k\,E_{lj} -\i\,\gd{}^i_l\,E_{kj}
                           +\ft{\i}{4}\gd{}^i_j\,E_{kl} \,,\non[3ex]
\Com{G^i{}_j}{F^{kl}} &=&  \i\,\gd{}^k_j\,F^{il} -\i\,\gd{}^l_j\,F^{ik}
                           -\ft{\i}{4}\gd{}^i_j\,F^{kl} \,,\non
\Com{G^i{}_j}{F_{kl}} &=&  \i\,\gd{}^i_k\,F_{lj} -\i\,\gd{}^i_l\,F_{kj}
                           +\ft{\i}{4}\gd{}^i_j\,F_{kl} \,.
\]
The remaining part of \EE acts as
\[
\Com{G^{ijkl}}{E_{mn}} &=&
 -\i\,\gd{}^{[ij}_{mn}\,E^{kl]}_{\vphantom{mn}}\,,\non[1ex]
\Com{G^{ijkl}}{E^{mn}} &=&
 -\ft{\i}{24}\,\epsilon^{ijklmnpq}\,E_{pq} \,, \non[1ex]
\Com{G^{ijkl}}{F_{mn}} &=&
 -\i\,\gd{}^{[ij}_{mn}\,F^{kl]}_{\vphantom{mn}} \,,\non[1ex]
\Com{G^{ijkl}}{F^{mn}} &=&
 -\ft{\i}{24}\,\epsilon^{ijklmnpq}\,F_{pq} \,.
\]
The grading of the generators is given by the dilatation generator
$H$
\[
\begin{array}{rclcrcl}
\Com{H}{E}      &=& -2\,\i\,E\,, &&
\Com{H}{F}      &=&  2\,\i\,F\,, \\[1ex]
\Com{H}{E^{ij}} &=& -\i\,E^{ij}\,, &&
\Com{H}{F^{ij}} &=&  \i\,F^{ij}\,, \\[1ex]
\Com{H}{E_{ij}} &=& -\i\,E_{ij}\,, &&
\Com{H}{F_{ij}} &=&  \i\,F_{ij}\,.
\end{array}
\]
The remaining non-vanishing commutation relations are
\[
\begin{array}{rclrcl}
\Com{E^{ij}}{F^{kl}} &=& 12\,\i\,G^{ijkl}\,, &
\Com{E^{ij}}{F_{kl}} &=& 4\,\i\,\delta^{[i}_{[k}\, G^{j]}{}_{l]}
                            -\i\,\delta_{kl}^{ij}\, H \,, \\[1ex]
\Com{E_{ij}}{F_{kl}} &=& -12\,\i\,G_{ijkl}\,, &
\Com{E_{ij}}{F^{kl}} &=&  4\,\i\,\delta^{[k}_{[i}\, G^{l]}{}_{j]}
                            +\i\,\delta^{ij}_{kl}\, H \,,\\[1ex]
%
%
\Com{E^{ij}}{E_{kl}} &=& 2\,\i\,\delta^{ij}_{kl} E\,, &
\Com{F^{ij}}{F_{kl}} &=& 2\,\i\,\delta^{ij}_{kl} F\,, \\[1ex]
\Com{E}{F^{ij}} &=& -\i\,E^{ij}\,, &
\Com{F}{E^{ij}} &=&  \i\,F^{ij}\,, \\[1ex]
\Com{E}{F_{ij}} &=& -\i\,E_{ij}\,, &
\Com{F}{E_{ij}} &=&  \i\,F_{ij}\,, \\[1ex]
\Com{E}{F} &=& \i\,H\,. &&\\[1ex]
\end{array}
\]
The \E commutation relations are the same as in \ct{GuKoNi00}, except
that the structure constants carry an extra factor of $\i$. As can
be verified by computation of the Cartan Killing form from the
structure constants that can be extracted from the above commutation
relations, the maximal compact subgroup \SO[16] is generated
by the following linear combinations of \E generators
\[
G^-_{ij}\,,\;\; G^-_{ijkl} \,,\;\;
E_{ij}+F^{ij}\,,\;\; E^{ij}-F_{ij}\,,\;\; E+F\,.
\]
This confirms that we are indeed dealing the split real form \E.

The quadratic \E Casimir operator (in a convenient normalization)
is a sum of three terms
\[
\lb{E8Casimir}
\cC_2\big[\E\big] = \cC_2\big[SL(2,\Rn)\big] + \cC_2\big[\EE\big] + \cC_2'
\]
with
\[
\cC_2\big[SL(2,\Rn)\big] &:=& \ft12 (EF +FE - \ft12 H^2) \,,\non
\cC_2\big[\EE\big] &:=& -\ft12\, G^i{}_j G^j{}_i -3\,G^{ijkl} G_{ijkl} \,,\non
\cC_2' &=& \ft14 \left( E^{ij} F_{ij} + F_{ij} E^{ij}
                       - E_{ij} F^{ij} - F^{ij} E_{ij} \right) \,.
\]
Here the terms in the first and second line represent the Casimir
operators of the $SL(2,\Rn)$ and \EE subalgebras, respectively.
When substituting the explicit expressions in terms of coordinates
and momenta for the generators, we obtain
\[
\cC_2\big[SL(2,\Rn)\big] &=& I_4 - \ft{3}{16}  \,, \non
\cC_2\big[\EE\big] &=& 3\,I_4 - \ft{969}{16}   \,,  \non
\cC_2' &=& -4\,I_4 - \ft{237}4 \,.
\]
Hence all terms containing the operators $X^{ij} , P_{ij}, y$
and $p$ actually cancel, leaving us with a constant value
for the minimal representation
\[
\cC_2\big[\E\big] = -120
\]
which is the \E analog of the result
\[
\cC_2\big[SL(2,\Rn)\big] = \ft14 g -\ft3{16} \,,
\]
familiar from conformal quantum mechanics -- except that there is no coupling
constant any more for \E that we can tune! Of course, unlike for the group
$SL(2,\Rn)$~\footnote{ Modulo the subtlety that the two singleton irreps of
  $SL(2,\Rn)$ have the same eigenvalue of the Casimir.}, this result does not
by itself imply the irreducibility of the minimal representation
of \E. To show that, we have to compute in addition the eigenvalues
of the higher order \E Casimir invariants. However, it is almost
self-evident that these, too, will collapse to $c$-numbers when
the coordinates and momenta are substituted, for the simple reason
that we cannot build \E invariants from the coordinate and momentum
operators alone. This is in stark contrast to the singleton representation of
\EE for which the coordinates and momenta do form a non-trivial (linear)
representation of \EE, permitting the construction of non-vanishing higher
order \EE invariants. This provides an independent argument that the minimal
representation of \E is indeed irreducible.

We conclude this section by giving the Chevalley generators
corresponding to the eight simple roots of \E, with the labeling
indicated in the figure.
\begin{figure}[h]
\begin{center}
\epsfig{file=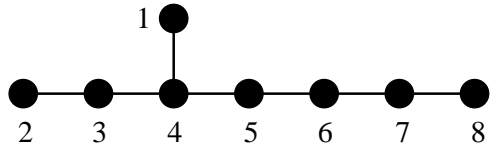}
\end{center}
\caption{Numbering of simple roots of \E}
\end{figure}

\noindent
They are
\[
\begin{array}{rclrcl}
e_1 &=& 12\,G^{4567}\quad , \quad & f_1 &=& 12\,G^{1238}\,, \\[1ex]
e_2 &=& G^1{}_2 \quad , \quad    & f_2 &=& G^2{}_1\,, \\
e_3 &=& G^2{}_3 \quad , \quad     & f_3 &=& G^3{}_2\,, \\[1ex]
e_4 &=& G^3{}_4 \quad , \quad     & f_4 &=& G^4{}_3\,, \\[1ex]
e_5 &=& G^4{}_5 \quad , \quad     & f_5 &=& G^5{}_4\,, \\[1ex]
e_6 &=& G^5{}_6 \quad , \quad      & f_6 &=& G^6{}_5\,, \\[1ex]
e_7 &=& G^6{}_7 \quad , \quad      & f_7 &=& G^7{}_6\,, \\[1ex]
e_8 &=& E^{78}  \quad , \quad      & f_8 &=& F_{78}\,.  \\[1ex]
\end{array}
\]
The generators of the Cartan subalgebra are given by
\[
\begin{array}{rclrcl}
h_1 &=& \multicolumn{2}{l}{G^4{}_4+G^5{}_5 +G^6{}_6+G^7{}_7}
    &=& -G^1{}_1 -G^2{}_2 -G^3{}_3 - G^8{}_8 \,,\\[1ex]
h_2 &=& G^1{}_1-G^2{}_2 \,,\\
h_3 &=& G^2{}_2-G^3{}_3 \,,\\[1ex]
h_4 &=& G^3{}_3-G^4{}_4 \,,\\[1ex]
h_5 &=& G^4{}_4-G^5{}_5 \,,\\[1ex]
h_6 &=& G^5{}_5-G^6{}_6 \,,\\[1ex]
h_7 &=& G^6{}_6-G^7{}_7 \,,\\[1ex]
h_8 &=& G^7{}_7+G^8{}_8-\ft12 H \,.
\end{array}
\]
Comparison with the formulas given in the appendix of \ct{KaPiWa01}
shows that the basis of coordinate vs. momentum variables used there
differs from ours by the choice of polarization (or ``Fourier
transformation''). Accordingly, the linearly realized
subgroup exposed there is also different from ours.

\mathversion{bold}
\section {Oscillator Representation}
\mathversion{normal}

In the coordinate representation, the linearly realized $SL(8,\Rn)$ subalgebra
of \EE plays a distinguished role. As is well known, however, there is another
basis of \EE, which is equally important in supergravity and superstring
theory, where $SL(8,\Rn)$ is replaced by $SU(8)$\ct{CreJul79}. In this
section, we demonstrate that the full \E Lie algebra can be rewritten in terms
of this complex basis.  The change of basis is equivalent to the replacement
of coordinates and momenta by creation and annihilation operators, and thus to
the replacement of the coordinate representation by a holomorphic
(Bargmann-Fock) representation. This will allow us to establish the connection
with the oscillator realization of \EE discovered already some time
ago\ct{GunSac82}. For this purpose, we introduce the creation and annihilation
(or raising and lowering) operators
\[
a^{AB} &:=& \ft1{4\sqrt{2}} \Gamma^{ij}_{AB} \big( X^{ij} - \i\,P_{ij} \big)
            \non
a_{AB} &:=& \ft1{4\sqrt{2}} \Gamma^{ij}_{AB} \big( X^{ij} + \i\,P_{ij} \big)
             \equiv \left( a^{AB}\right)^\dag
\]
The normalization has been chosen such that
\[
\big[\,  a_{AB}\, ,\, a^{CD} \, \big] = \delta_{AB}^{CD}
\]
Substituting these operators into (\ref{e7gen}) and defining
\[
G^A{}_B  &:=& - \ft14 \i\,G^{-}_{ij}\Gamma^{ij}_{AB}
             +\ft18 \,G^{-}_{ijkl}\Gamma^{ijkl}_{AB} = G_B{}^A ,\non
G^{ABCD} &:=& \left( \ft1{48} \i\,G^{+}_{ik}\delta_{lj}
    - \ft1{16} G^{+}_{ijkl}\right) \Gamma^{ij}_{[AB}\Gamma^{kl}_{CD]}\,,
\]
we obtain the singleton representation of \EE in the $SU(8)$ basis
of\ct{GunSac82}
\[
G^A{}_B  &:=& 2 a^{AC} a_{BC} - \ft14 \delta^A_B a^{CD} a_{CD} \non
G^{ABCD} &:=& \ft12 a^{[AB} a^{CD]} -
              \ft1{48} \epsilon^{ABCDEFGH} a_{EF} a_{GH}
\]
where $G^A{}_B$ now generates the $SU(8)$ subgroup of \EE. In
deriving this result, we made use of the formulas (see e.g. the
appendices of\ct{CreJul79,deWNic86})
\[
\Gamma^{ijkl}_{AB} &=& - \ft1{24} \epsilon^{ijklmnpq}\,\Gamma^{mnpq}_{AB}
\]
and
\[
\Gamma^{ij}_{[AB} \Gamma^{kl}_{CD]} &=&
  -\ft23 \delta\oversym{^{k[i} \Gamma^{j]m}_{[AB}
          \Gamma^{ml}_{CD]}\!\!\!}\,\,\, +
   \Gamma^{[ij}_{[AB} \Gamma^{kl]}_{CD]}   \non
\Gamma^{ik}_{[AB} \Gamma^{kj}_{CD]} &=&
\ft1{24} \epsilon_{ABCDEFGH} \Gamma^{ik}_{EF} \Gamma^{kj}_{GH}  \non
\Gamma^{[ij}_{[AB} \Gamma^{kl]}_{CD]} &=&
 - \ft1{24} \epsilon_{ABCDEFGH} \Gamma^{[ij}_{EF} \Gamma^{kl]}_{GH} \non
&=& + \ft1{24} \epsilon^{ijklmnpq}\,\Gamma^{mn}_{[AB} \Gamma^{pq}_{CD]}
\]
We note the complex (anti)self-duality relation
\[
G_{ABCD} \equiv \left( G^{ABCD} \right)^\dag
        =  - \ft1{24} \epsilon_{ABCDEFGH} G^{EFGH}
\]
and the commutation relations
\[
\Com{G^A{}_B}{G^{C\vphantom{D}}{}_D} &=&
  \gd{}_{B}^{C}\, G^A{}_{D} -\gd{}_{D}^{A}\, G^C{}_{B} \,,\non[2ex]
\Com{G^A{}_B}{G^{CDEF}}  &=& -4\,\gd{}_{B}^{[C}\, G^{DEF]A}_{\phantom{B}}
                   -\ft{1}{2}\, \gd{}^A_B G^{CDEF}_{\phantom{B}}\,,\non[2ex]
\Com{G^{ABCD}}{G^{EFGH}} &=& \ft{1}{36}\,\epsilon^{ABCDI[EFG}\,
                                 G^{H]}{}_{I} \,.
\]
In order to render the remaining \E generators \SU[8] covariant,
we define
\[
\begin{array}{rclcrcl}
E^{AB} &:=&
  \ft1{4\sqrt{2}} \Gamma^{ij}_{AB} \big(E^{ij}-\i\,E_{ij}\big)
         = y a^{AB}\,,\\[1ex]
E_{AB} &:=&
  \ft1{4\sqrt{2}} \Gamma^{ij}_{AB} \big(E^{ij}+\i\,E_{ij}\big)
         = y a_{AB}\,,\\[1ex]
\end{array}
\]
The computation is more tedious for the generators which are cubic
and quartic in the oscillators, and most conveniently done by
checking the \E algebra again. We have
\[
F &=& \ft12 p^2 + 2\,y^{-2} I_4(a,a^\dag)
\]
with the \SU[8] invariant expression for $I_4$ in terms of oscillators
\[
I_4(a,a^\dag)
&\=& I_4 (X,P)\non
&=&  +\ft12(a^{AB}a_{BC}a^{CD}a_{DA}+a_{AB}a^{BC}a_{CD}a^{DA})\non
&&   -\ft18(a^{AB}a_{AB}a^{CD}a_{CD}+a_{AB}a^{AB}a_{CD}a^{CD})\non
&&   +\ft{1}{96}\,\epsilon^{ABCDEFGH} a_{AB}a_{CD}a_{EF}a_{GH} \non
&&   +\ft{1}{96}\,\epsilon_{ABCDEFGH} a^{AB}a^{CD}a^{EF}a^{GH}
     +\ft{547}{16} \,.
\]
Note that the normal-ordered version of $I_4$ (with all the annihilators to
the right) is {\it not} \EE invariant, because
\[
I_4(a,a^\dag) \;=\; \; :\!I_4(a,a^\dag)\!: \, -\ft{\i}{4} \cN - 49
\]
with the number operator
\[
\cN := a^{AB} a_{AB}
\]
which does not commute with $G^{ABCD}$.

The remaining generators are now straightforwardly deduced by
commuting $F$ with $E^{AB}$ and $E_{AB}$:
\[
F^{AB} &:=&
  \ft1{4\sqrt{2}} \Gamma^{ij}_{AB} \big(F^{ij}-\i\,F_{ij} \big)
\;\equiv\; \i\,\Com{E^{AB}}{F} \non
&=& -p\, a^{AB}
   +\ft{\i}{2} y^{-1}(a^{AB} a_{CD} a^{CD} + a^{CD} a_{CD} a^{AB}) \non
&& + 4\i y^{-1}\, a\,\oversym{\!{}^{AC} a_{CD} a^{DB}\!}\,
   -\ft{\i}{12} y^{-1} \epsilon^{ABCDEFGH} a_{CD} a_{EF} a_{GH} \,,\non
F_{AB} &:=&
  \ft1{4\sqrt{2}} \Gamma^{ij}_{AB} \big(F^{ij}+\i\,F_{ij} \big)
\;\equiv\; \i\,\Com{E_{AB}}{F} \non
&=& -p\, a_{AB}
   -\ft{\i}{2} y^{-1} (a_{AB} a^{CD} a_{CD} + a_{CD} a^{CD} a_{AB})  \non
&& - 4\i y^{-1}\, a\,\undersym{\!{}_{AC} a^{CD} a_{DB}\!}\,
   +\ft{\i}{12} y^{-1}\epsilon_{ABCDEFGH} a^{CD} a^{EF} a^{GH} \,.
\]
These generators now transform covariantly under the \SU[8] group
\[
\Com{G^A{}_B}{E_{CD}} &=&
  \gd{}^A_C\,E_{DB} -\gd{}^A_D\,E_{CB}
 +\ft{1}{4}\gd{}^A_B\,E_{CD} \,,\non[3ex]
\Com{G^A{}_B}{F_{CD}} &=&
  \gd{}^A_C\,F_{DB} -\gd{}^A_D\,F_{CB}
 +\ft{1}{4}\gd{}^A_B\,F_{CD} \,,
\]
and the remaining part of \EE acts as
\[
\Com{G^{ABCD}}{E_{EF}} &=&
  -\gd{}^{[AB}_{EF}\,E^{CD]}_{\vphantom{EF}}\,,\non
\Com{G^{ABCD}}{F_{EF}} &=&
  -\gd{}^{[AB}_{EF}\,F^{CD]}_{\vphantom{EF}}\,.
\]
Furthermore,
\[
\Com{E_{AB}}{E^{CD}} &=& 2 \delta_{AB}^{CD}\,E \,, \non
\Com{F_{AB}}{F^{CD}} &=& 2
\delta_{AB}^{CD}\,F \,, \non
\Com{E^{AB}}{F^{CD}} &=& -12\,\i\,G^{ABCD} \,,\non
\Com{E_{AB}}{F^{CD}} &=&
  -2\,\i\,\delta_{[A}^{[C} G^{D]}_{\vphantom{[]}}{}_{B]}
  -\delta_{AB}^{CD}\, H  \,.
\]

It remains to discuss the $SL(2,\Rn)$ subgroup, for which we
likewise switch to a complex basis (the $SU(1,1)$ basis)
\[\label{su11}
L_0 &:=& \ft12(E+F) = \ft14 ( p^2 + y^2) + y^{-2} I_4(X,P)   \,,\non
L_{1} &:=& \ft12(E-F+\i H) =\ft14 (y+\i p)^2 - y^{-2} I_4(X,P) \,,\non
L_{-1} &:=& \ft12(E-F-\i H) =\ft14 (y-\i p)^2 - y^{-2} I_4(X,P) \,,
\]
In the absence of the term containing $y^{-2}$, it would again be
convenient to employ creation and annihilation operators
$b:=\ft1{\sqrt{2}}(y+ip)$ and $b^\dag = \ft1{\sqrt{2}}(y-ip)$
to recover the well known singleton representation of $SU(1,1)$,
but for non-vanishing value of the quartic \EE invariant
it is not possible to switch this term off. The presence
of $y^{-1}$ and $y^{-2}$ makes it somewhat awkward to express
all generators in terms of creation and annihilation operators,
so one might prefer to keep the coordinate representation in
this sector. The commutation relations are, however, not affected
by the choice of variables.
\[
\Com{L_0}{L_{\pm 1}} &=& \mp L_{\pm 1}\,,\qquad
\Com{L_{+1}}{L_{-1}} \;=\; 2\, L_{0}\,.
\]
This basis is no longer hermitean, but
\[
L_0^\dag = L_0 \; , \quad (L_{\pm 1} )^\dag = L_{\mp 1}
\]

Diagonalizing the new hermitean generator $2 L_0$ instead of $H$
we obtain an alternative 5-graded decomposition
\[
\fe_8 &=&   \fk^{-2} \oplus \fk^{-1} \oplus \fk^0 \oplus \fk^{+1}
  \oplus \fk^{+2} \,.  \label{5grading-new}
\]
such that
\[
(\fk^0)^\dag = \fk^0 \; , \quad
(\fk^{\pm 1})^\dag = \fk^{\mp 1} \; , \quad
(\fk^{\pm 2})^\dag = \fk^{\mp 2}
\]
We note that, strictly speaking, the elements of $\fk^n$ by themselves
do not belong to the real Lie algebra $\fe_8$ as defined in section 2,
but to its complexification. It is only the hermitean linear combinations
which do.

The subspaces of grade $\pm2$ are one-dimensional with generators $L_{\mp1}$,
the grade 0 space is spanned by the \EE generators\Eq{e7gen} together with
$L_0$ and the subspaces of grade $\pm1$ are generated by $\cE^{AB},\cE_{AB}$
and $\cF^{AB},\cF_{AB}$ defined by
\[
\cE^{AB} &:=& \ft{1}{\sqrt2}(E^{AB}+\i\,F^{AB})\,,\non
\cE_{AB} &:=& \ft{1}{\sqrt2}(E_{AB}+\i\,F_{AB})\,,\non
\cF^{AB} &:=& \ft{1}{\sqrt2}(E^{AB}-\i\,F^{AB})\,,\non
\cF_{AB} &:=& \ft{1}{\sqrt2}(E_{AB}-\i\,F_{AB})\,,
\]
respectively. In terms of these generators all structure constants
are real:
\[
\Com{L_{+1}}{\cE^{AB}} &=& 2\,\cF^{AB} \,, \non
\Com{L_{-1}}{\cF^{AB}} &=& -2\,\cE^{AB} \,, \non
\Com{\cE_{AB}}{\cE^{CD}} &=& 2\,L_{-1} \,, \non
\Com{\cF_{AB}}{\cF^{CD}} &=& 2\,L_{+1} \,, \non
\Com{\cE^{AB}}{\cF^{CD}} &=& -12\,G^{ABCD} \,,\non
\Com{\cE_{AB}}{\cF^{CD}} &=&
  -2\,\delta_{[A}^{[C} G^{D]}_{\vphantom{[]}}{}_{B]}
  +2\,\delta_{AB}^{CD}\, L_0  \,.
\]

\mathversion{bold}
\section{Decompositions and Truncations}
\mathversion{normal}

\mathversion{bold}
\subsection {$\EE\times SL(2,\Rn)$ decomposition of the minimal unitary
  representation of \E}
\mathversion{normal}

A non-compact group $G$ admits unitary representations of the lowest weight
(or highest weight) type if and only if the quotient $G/K$ of $G$ with respect
to its maximal compact subgroup $K$ is an Hermitean symmetric
space\ct{Hari56,Hari56a}. From this theorem it follows that the simple
non-compact groups that admit lowest (highest) weight unitary representations
are \SO[n,2], \SU[n,m], \SOS[2n], \SpR[2n], \6[(-14)], and \7[(-24)]. The
unitary lowest (highest) weight representations belong to the holomorphic
(anti-holomorphic) discrete series and within these representations the
spectrum of, at least, one generator is bounded from below (above). More
generally, a non-compact group $G$ admits representations belonging to the
discrete series if it has the same rank as its maximal compact
subgroup\footnote{For an excellent introduction the general theory of
unitary representations of non-compact groups see\ct{Knapp}.}. Thus the
non-compact group \E as well as \EE admit discrete series representations.
However, they are not of the lowest or highest weight type.

In this subsection we will analyze the decomposition of the minimal
representation of \E with respect to its subgroup $\EE\times SL(2,\Rn)$,
but using the complex basis of the last section. The group \SU[1,1]
admits holomorphic and anti-holomorphic unitary representations,
and they exhaust the list of discrete series representations
for \SU[1,1]. (This is not true for higher rank non-compact groups
admitting such representations.) As mentioned earlier, the realization
of the \SU[1,1] subgroup within the minimal unitary representation
of \E is precisely of the form that arises in conformal
quantum mechanics\ct{AlFuFu76}. This is perhaps not surprising since
we obtained our realization from the geometric action of \E as
a quasi-conformal group in 57 dimensions\ct{GuKoNi00}.

By comparison of the \SU[1,1] subgroup\Eq{su11} with that of\ct{AlFuFu76}
it follows that the coupling constant $g$ in conformal quantum
mechanics is simply
\[
g &=& 4\, I_4(X,P)
\]
in our realization. The quadratic Casimir of \SU[1,1] is
\[
\cC_2 \big[\SU[1,1]\big]
&=& L_0^2 -\ft12\, (L_{1}L_{-1}+L_{-1}L_{1}) = I_4(X,P) - \ft{3}{16} \,.
\]
Thus for a given eigenvalue of $I_4(X,P)$, we are led to a
definite unitary realization of \SU[1,1]. As we showed above,
$I_4(X,P)$ is simply  the quadratic Casimir operator of \EE
(cf.\Eq{e7-casimir}) that commutes with \SU[1,1], up to an
additive normal ordering constant. Hence classifying all the
possible eigenvalues of $I_4(X,P)$ within the minimal unitary
realization of \E is equivalent to giving the decomposition of the
\E representation with respect to
$\EE\times\SU[1,1]$\footnote{Here we should note that  in certain
exceptional cases the eigenvalue of the quadratic Casimir operator
may not uniquely label the unitary irreducible representation of
\SU[1,1]. In such cases one needs to use additional labels such as
the eigenvalues of $L_0$.}. Unitarity requires the eigenvalues of
the Casimir operators to be real. Therefore all the eigenvalues of
$I_4$ must be real. As we showed above, the realization of \EE
within \E coincides with the singletonic oscillator realization of
\EE\ct{GunSac82}. The oscillator realization of  \EE leads to an
infinitely reducible unitary representation\ct{GunSac82}. Hence
the minimal representation of \E will be infinitely reducible with
respect to $\EE\times\SU[1,1]$.

Denoting the eigenvalues of the quadratic Casimir operator \SU[1,1] as
\[
\cC_2\big[ \SU[1,1]\big] = j(j-1)
\]
we find that
\[
j &=& \ft12 \pm\sqrt{I_4 +\ft1{16}}
\]
Depending on the eigenvalue of $I_4$ we will be led to one of the
well-known series of representations of \SU[1,1]\ct{Barg,Knapp}
\begin{itemize}
\item[(a)] Continuous principal series:
\[
j &=& \ft12 - \i\,\rho\,, \qquad 0<\rho<\infty\,, \non
j(j-1) &=& -(\ft14 + \rho^2) \;<\; -\ft14
\]
with the eigenvalues $\ell$ of $L_0$ unbounded from above and from below
\[
\ell &=& \ell_0,\, \ell_0\pm1,\, \ell_0\pm2\,, \ldots \qquad (\ell_0\in\Rn)
\]
\item[(b)] Continuous supplementary series:
\[
 |j-\ft12| &<& \ft12 - |\ell_0|\,, \qquad (j,\ell_0\in\Rn)
\]
again with unbounded eigenvalues $\ell$ of $L_0$ in both directions.
\item[(c)] Holomorphic discrete series $D^{+}(j)$ (lowest weight irreps):
\[
j &>& 0 \quad ,\quad \ell_0\;=\;j\,, \qquad (j\in\Rn) \non
\ell &=& j,\, j+1,\, j+2,\, \ldots
\]
\item[(c)] Discrete series $D^{-}(j)$ (highest weight irreps):
\[
j &>& 0 \quad , \quad \ell_0\;=\;-j\,, \qquad (j\in\Rn) \non
\ell &=& -j,\, -j-1,\, -j-2,\, \ldots
\]
\end{itemize}

For vanishing quartic invariant $I_4$ we find
\[
j&=&\ft14 \quad\text{or}\quad j\;=\;\ft34
\]
corresponding to the two singleton irreps of \SU[1,1]. Note that
the eigenvalues of the Casimir operator for the two singleton
irreps coincide. They are distinguished by the value of $j$ which
is the eigenvalue of $L_0$ on the corresponding lowest weight vector.

\mathversion{bold}
\subsection {The \SU[2,1] truncation of \E}
\mathversion{normal} To give the decomposition of the minimal
representation of \E w.r.t. its $ \EE \times SL(2,R)$ subgroup we
need to determine all possible eigenvalues of $I_4(X,P)$ within
our realization. Since this determination appears to be rather
complicated, we shall study this question in a somewhat simpler
setting by truncating \E  to a special subgroup. \E has the
subgroup $\6[(2)]\times\SU[2,1]$, where \6[(2)] has the maximal
compact subgroup $\SU[6]\times\SU[2]$. It is instructive to
truncate our realization to the \SU[2,1] subgroup, which is one of
the minimal subgroups that admit a non-trivial 5-grading. This is
achieved by introducing coordinate $x$ and momentum $p_x$
corresponding to the symplectic trace components of $X^{ij}$ and
$P_{ij}$, respectively:
\[
x   &:=& - \ft{\sqrt{2}}{4}\, \Omega_{ij} X^{ij} \,, \non
p_x &:=&   \ft{\sqrt{2}}{4}\, \Omega^{ij} P_{ij} \,,
\]
and throwing away the symplectic traceless components in our
realization. The matrix $\Omega_{ij}$ is the symplectic metric defined by
\[
\Omega_{ij} &:=&
\left(\begin{array}{cc}0&-\ID\\ \ID&0\end{array}\right)
\]
and $\Omega^{ij}$ is its inverse: $\Omega_{ij}\Omega^{jk}=\delta_i^k$.
The generators $x$ and $p_x$ obey the canonical commutation relation
\[
 \Com{x}{p_x} &=&\i \,.
\]

The resulting expressions for the generators of \SU[2,1] are
\[
E &:=& \ft12\, y^2 \,,\non[2ex] E_{\uparrow}
  &:=& y\,x \,,\non[2ex] E_{\downarrow}
  &:=& y\,p_x \,,\non[2ex] H     &:=& \ft{1}{2} (y\,p_y+p_y\,y) \non[2ex]
A &:=&-\ft{3}{4} (x^2+p_x^2) \non[2ex]
F_{\uparrow}   &:=& -x p_y  -\ft{1}{2}y^{-1}(x p_x x +p_x^3) \,,\non[2ex]
F_{\downarrow} &:=& p_x p_y -\ft{1}{2}y^{-1}(p_x x p_x +x^3) \,,\non[2ex]
F &:=& \ft12 p_y^2 +\ft{1}{8}y^{-2} (x^2+p_x^2)^2 -\ft{1}{8}y^{-2}\,,
\]

The 5-grading in the above basis is with respect to $H$ which is
the non-compact dilatation generator. To understand the resulting
representation of \SU[2,1] it is useful to go to the basis in which
the Lie algebra of \SU[2,1] has a 3-grading w.r.t. the compact
$U(1)$ generator that commutes with the \SU[2] subgroup.
\[
\SU[2,1] &=& \left(\begin{array}{c}L_{-}\\K_{-}\end{array}\right) \oplus
\left(\begin{array}{c}J_{i}\\J_{0}\end{array}\right) \oplus
\left(\begin{array}{c}L_{+}\\K_{+}\end{array}\right)
\]
The compact $U(1)$ generator is
\[
J_{0} &=& \ft12 (E+F-\ft23\,A) \;=\; L_{0}-\ft13\,A
\]
and the \SU[2] generators are
\[
J_1 &=& \ft1{2\sqrt{2}} (E_{\uparrow}+F_{\downarrow})\,, \non
J_2 &=& \ft1{2\sqrt{2}}
(E_{\downarrow}+F_{\uparrow})\,, \non J_3 &=& \ft14 (E+F+2\,A)\,,
\]
with the commutation relations
\[
\Com{J_i}{J_j} &=& \i\,\epsilon_{ijk} J_{k}\,,
\quad (i,j,k=1,2,3) \non \Com{J_0}{J_i}
&=& 0\,.
\]
The grade $\pm1$ generators are
\[
K_{\pm } &=& (E_{\uparrow}-F_{\downarrow})
      \mp \i(E_{\downarrow}-F_{\uparrow}) \,,\non
L_{\pm } &=& \ft12(E-F\mp\i H) \,,
\]
satisfying
\[
\Com{J_0}{K_{\pm }} &=& \pm K_{\pm }\,,\non
\Com{J_0}{L_{\pm }} &=& \pm L_{\pm }\,.
\]
More explicitly we have
\[
K_{\pm } &=& \{(y\mp ip_y) + \ft{1}{2}y^{-1} [1+(x^2+p_x^2)] \}
              (x\mp ip_x) \, ,\non
L_{\pm }&=& \{ \ft{1}{4} (y\mp ip_y)^2 -
             \ft{1}{16}y^{-2} [-1 + (x^2+p_x^2)^2] \}
\]

The generator $J_0$ that determines the 3-grading of $SU(2,1)$ w.r.t.
maximal compact subgroup $SU(2)\times U(1)$ is manifestly positive
definite. In terms of the annihilation and creation operators
$a_x, a_x^{\dagger}$ the generator $J_0$ takes the form:
\[
J_0=\ft14 \{ y^2 + p_y^2 + y^{-2} N_x(N_x+1) +2N_x+1 \}
\]
where $N_x= a_x^{\dagger}a_x $ is the number operator. This implies
that the resulting unitary irreducible representations of $SU(2,1)$
must be of the lowest weight type (positive energy). Now the quadratic
Casimir operator of $SU(2,1)$  is given by
\[
\cC_2\big[SU(2,1)\big] &=& \ft12 (EF+FE) -\ft14 H^2 + \ft13 A^2 \non
    && + \ft14 (E_{\uparrow}F_{\downarrow} +F_{\downarrow}
    E_{\uparrow}) +\ft14 (E_{\downarrow}F_{\uparrow} +
    F_{\uparrow} E_{\downarrow})
\]
Substituting the expressions for the generators we find that the the
quadratic Casimir of $SU(2,1)$ becomes simply a $c$-number, namely
\[
\cC_2\big[SU(2,1)\big] = -\ft{3}{16}
\]
which suggests that the representation may be irreducible. To have an
irreducible representation the cubic Casimir must likewise reduce to a
$c$-number. However, we can study the irreducibility of the representation
without having to calculate the cubic Casimir. Since we know that
our representation is of the lowest weight type we can use the fact
that a unitary irrep of the lowest weight type is uniquely determined
by a set of states $| \Omega \rangle$ transforming irreducibly under
the maximal compact subgroup $SU(2) \times U(1)$ and that are annihilated
by all the generators of grade $-1$ under the 3-grading. Thus we need
to find all such states that are annihilated by $K_{-}$ and $L_{-}$:
\[
K_{-}|\Omega \rangle &=& 2 \Big( a_y + \ft{1}{\sqrt{2}} y^{-1}
    (N_x+1)\Big) a_x |\Omega\rangle = 0 \non
L_{-}   |\Omega \rangle &=& \ft12 \Big(a_y^2 - \ft12
y^{-2}N_x(N_x+1)
       \Big)|\Omega \rangle =0
\]
and that transform irreducibly under the maximal compact subgroup
$SU(2) \times U(1)$.
 The only normalizable state satisfying these conditions is
the one particle excited state
\[
a_y^{\dagger} |0 \rangle
\]
 where the vacuum state $|0 \rangle $ is annihilated by both annihilation
operators
\[
a_x|0 \rangle =0 \non a_y |0 \rangle =0
\]

This state is a singlet of $SU(2)$ and has the $U(1)$ charge 1.
This proves that the minimal realization of $SU(2,1)$ obtained by
truncation of the minimal representation of \E is also
irreducible.

The subgroup  $\EE \times SU(1,1)$ of \E under truncation to
$SU(2,1)$ reduces to $U(1) \times SU(1,1)$.
 The generators of $SU(1,1)$ subgroup of $SU(2,1)$ are
given by
\[
L_{+}&=& \ft{1}{2} (E-F-iH) \non &=&
 \{ \ft{1}{4} (y- ip_y)^2 -
     \ft{1}{16} y^{-2} [-1 + (x^2+p_x^2)^2] \} \non
L_{-}&=& \ft{1}{2} (E-F+iH) \non &=& \{ \ft{1}{4} (y + ip_y)^2 -
     \ft{1}{16} y^{-2} [-1 + (x^2+p_x^2)^2] \} \non
 L_0 &=& \ft{1}{2} (E+F) \non
&=&  \{ \ft14 (y^2 + p_y^2) + \ft{1}{16}y^{-2} [-1 +
(x^2+p_x^2)^2] \}
\]
They are therefore the analogs of the generators(\ref{su11}). The
generator of  $U(1)$ that commutes with $SU(1,1)$ is simply $A$.
In this truncation the quartic \EE invariant reduces to
\[
I_4= \ft{1}{16}[(x^2+p_x^2)^2-1]
\]
which can be written in terms of the annihilation and creation
operators as
\[
I_4= \ft{1}{4} \{ (a_x^{\dagger}a_x) (a_x^{\dagger}a_x +1) \}
\]
with the (obvious) identifications $a_x=-\ft{\sqrt{2}}4
\Omega^{AB} a_{AB}$ and $a_x^\dag=\ft{\sqrt{2}}4 \Omega_{AB}
a^{AB}$. Since the eigenvalues of the number operator
$a^{\dagger}a$ are non-negative integers n, the eigenvalues of
$I_4$ are simply $\ft14 n(n+1)$. In fact the realization of
$SU(1,1)$ in this case  leads to unitary lowest weight
representations of the type studied in \cite{AlFuFu76}. For $n=0$
we get the singleton irreps of $SU(1,1)$ corresponding to the
values $j=\ft14$ and $\ft34$. Thus the minimal unitary
representation of $SU(2,1)$ decomposes into a discretely infinite
set of irreps  of $SU(1,1)$ labelled by the eigenvalues of the
$U(1)$ generator which is the analog of the \EE subgroup of \E for
the $SU(2,1)$ truncation.

\section{Outlook}
The finite dimensional conformal group $SU(1,1)$ is well known
to possess an infinite dimensional extension (the Witt-Virasoro
group). One may therefore ask whether there exists a generalization
of this fact to \E. In other words, does there exist an infinite
dimensional Lie algebra (or Lie superalgebra) that contains the
Witt-Virasoro algebra and \E at the same time? While there appears
to be no linear Lie algebra with this property (and no finite
dimensional Lie superalgebra, either), an infinite dimensional
non-linear algebra of ${\cal W}$-type does exist. It is a nonlinear
quasi-superconformal algebra denoted as $QE_{8(8)}$ \cite{BinGun}.
The quasi-superconformal algebras in two dimensions were first
introduced in\cite{Poly} and further systematized in\cite{Bers}.
They were generalized in\cite{Roma} where two infinite families of
nonlinear quasi-superconformal algebras were introduced. A classification
of complex forms of quasi-superconformal algebras was given in \cite{FraLin}.
In \cite{BinGun} a complete classification and a unified realization of
the real forms of quasi-superconformal algebras were given.

In the infinite central charge limit the exceptional quasi-superconformal
algebra $QE_{8(8)}$ has the Lie algebra \E as a maximal finite dimensional
simple Lie subalgebra. The realization of $QE_{8(8)}$ given in\cite{BinGun}
involves 56 dimension $1/2$ bosons and a dilaton, which leads to a
realization of \E in the infinite central charge limit. It will be
important to understand if the resulting realization can be related
to the one given here. Furthermore one would like to know if one could
use the methods of \cite{BinGun} to give a unified realization of the
unitary representations of non-compact groups that act as
quasi-conformal groups as formulated in \cite{GuKoNi00}, thus
generalizing our minimal realization \E to all such noncompact groups.

\bigskip
\noindent {\bf Acknowledgments:} We are grateful to B.~Pioline and
A.~Waldron for discussions, and for sending us an advance copy of
ref.\ct{KaPiWa01}. We would also like to thank R. ~Brylinski for
bringing the references \cite{Vogan81} and \cite{GroWal94} to our
attention and for  helpful discussions.

\end{document}